\documentstyle[11pt]{article}

\def\thefootnote{\fnsymbol{footnote}}

\newcommand{\eq}{\begin{equation}}
\newcommand{\en}{\end{equation}}
\newcommand{\eqa}{\begin{eqnarray}}
\newcommand{\ena}{\end{eqnarray}}

\newcommand{\NP}[1]{Nucl.\ Phys.\ {\bf #1}}
\newcommand{\PL}[1]{Phys.\ Lett.\ {\bf #1}}

\newcommand{\PR}[1]{Phys.\ Rev.\ {\bf #1}}
\newcommand{\PRL}[1]{Phys.\ Rev.\ Lett.\ {\bf #1}}

\newcommand{\IJMP}[1]{Int.\ J.\ Mod.\ Phys.\ {\bf #1}}

\begin{document}
\begin{titlepage}

\vskip0.5cm
\begin{flushright}
DFTT 57/93\\
CBPF--NF 71/93\\
November 1993
\end{flushright}
\vskip0.5cm

\begin{center}
{\Large\bf   On the behaviour of spatial Wilson loops}
\vskip0.2cm
{\Large\bf in the high temperature phase of L.G.T.}
\end{center}

\vskip 1.3cm
\centerline{ M. Caselle$^a$, R.Fiore$^b$, F. Gliozzi$^a$,
P. Guaita$^c$  and  S. Vinti$^{a,d}$}
\vskip .6cm
\centerline{\sl  $^a$ Dipartimento di Fisica
Teorica dell'Universit\`a di Torino}
\centerline{\sl Istituto Nazionale di Fisica Nucleare, Sezione di Torino}
\centerline{\sl via P.Giuria 1, I-10125 Torino, Italy
\footnote{e--mail: caselle, gliozzi, vinti~@to.infn.it}}
\vskip .4 cm
\centerline{\sl $^b$ Dipartimento di Fisica, Universit\`a della Calabria}
\centerline{\sl Istituto Nazionale di Fisica Nucleare, Gruppo collegato di
Cosenza}
\centerline{\sl Arcavacata di Rende, I-87036 Cosenza, Italy
\footnote{e--mail: fiore~@cosenza.infn.it}}
\vskip .4 cm
\centerline{\sl  $^c$ Dipartimento di Fisica
Sperimentale dell'Universit\`a di Torino}
\centerline{\sl Istituto Nazionale di Fisica Nucleare, Sezione di Torino}
\centerline{\sl via P.Giuria 1, I-10125 Torino, Italy
\footnote{e--mail: guaita~@to.infn.it}}
\vskip .4 cm
\centerline{\sl $^d$ CBPF -- Centro Brasileiro de Pesquisas
Fisicas}
\centerline{\sl Rua Dr. Xavier Sigaud, 150}
\centerline{\sl 22290  Rio de Janeiro, Brazil}
\vskip 1.cm

\begin{abstract}
 The behaviour of the  space-like string tension in the high temperature
phase is studied. Data obtained in the $Z_2$ gauge model in (2+1)
dimensions are compared with predictions of a simple model of a
fluctuating flux tube with finite thickness.
It is shown that in the high temperature phase contributions coming
from the fluctuations of the flux tube vanish.
As a consequence we also show that in (2+1) dimensional
gauge theories the thickness of the flux tube coincides with
 the inverse of the deconfinement temperature.

\end{abstract}
\end{titlepage}

\setcounter{footnote}{0}
\def\thefootnote{\arabic{footnote}}

\section{Introduction}
\vskip .3cm

It is well known that in finite temperature Lattice Gauge Theories
(LGT) the ``spatial'' string
tension  (namely that extracted from Wilson loops orthogonal to the
compactified imaginary time direction) is no longer the order parameter
of confinement and is, in general, different from zero even in the
deconfined phase. Recently, in a set of interesting papers~\cite{t1,
klmpr,bfhks}  the behaviour of this spatial string tension was studied
for SU(2) and SU(3) gauge theories in (2+1) and (3+1) dimensions, leading
to new ideas and to a better understanding of the high temperature
behaviour of LGT's. At the same time, it was observed in~\cite{t1,t2} in the
case of SU(2) in (2+1) dimensions (and  it  will be confirmed in the present
paper, in the case of the gauge Ising model), that the inverse of the
 deconfinement temperature  $1/T^c$  almost coincides
 with the thickness $L_c$ of
the flux tube joining the quark-antiquark pair in the confined phase.
This implies that looking to space-like Wilson loops in the deconfined
phase  is almost equivalent to probing the interior of the flux tube
and gives us
a powerful tool to test the  effective flux tube models
of confinement. Moreover, the fact that $1/T^c$ and $L_c$ have
similar values is a quite interesting phenomenon in itself
and deserves further investigation.

In this paper we want to further pursue this
analysis, comparing some predictions of the effective flux tube picture
of LGT's with a set of high statistics Monte Carlo data on the 3d Ising
gauge model. The main results of our analysis are the following:

\begin{description}
\item{a]}
In the deconfined region ($T>T^c$) the space-like
 string tension increases as the temperature increases.
This trend is rather impressive: almost two orders of magnitude are gained
moving from the deconfinement point to the highest temperature
we can measure . Moreover in the deconfined phase
two distinct regimes can be identified: a first, rather
smooth, crossover region ($T^c<T<2T^c$) and a second, high temperature
regime ($T>2T^c$), where
 the string tension scales with the
temperature, which is the only remaining dimensional scale in this
regime.
In particular, this second region ($T>2T^c$) is the one in which
dimensional reduction has been shown to apply in the case of
the (3+1) dimensional SU(2) gauge theory~\cite{bfhks}.
All these results are in complete agreement with the scenarios
described in ref.s~\cite{t1,klmpr,bfhks}, and can be understood within
the framework of a ``compressed'' flux tube model.

\item{b]}
In the range $T^c~<~T~<~2T^c$
the  simple picture of a  compressed flux tube of uniform flux density
for the space-like
Wilson loops is quite accurate. It predicts a  string
tension rising linearly with the temperature and allows a rather
precise estimate of the thickness $L_c$ of the flux tube.

\item{c]}
Under the assumption that
in the deconfined region ($T>T^c$) the space-like flux tube
is ``frozen'' and the contribution of its transverse quantum
fluctuations is zero, it is possible to show that $L_c$ and $1/T^c$ must
coincide.
Suitable ratios of Wilson loops  can be introduced to
test this assumption, which turns out to be in good agreement with our
Monte Carlo simulations. These ratios can be used as order parameters for this
phase, being zero in the high temperature region  and
different from zero in the confining phase.

\item{d]}
There is an impressive agreement between some
dimensionless ratios of physical observables
that we obtain in the $Z_2$
case and those obtained by Teper~\cite{t1,t2} for SU(2).
This fact suggests
that the  behaviour of the theory near the deconfinement point is
dominated by the macroscopic properties of the flux tube, which are
largely independent of the precise
ultraviolet details of the models and only relies on the infrared
effective string action. As a consequence some ``super-universal''
behaviours, like those that we have observed, are expected for
dimensionless ratios of physical observables.
Assuming this point of view,
we expect that the effective models and ideas
that we shall discuss in the next section
should be valid for any gauge theory in 2+1 dimension.

\end{description}

Let us conclude this section by noticing that in (3+1) dimensions,
there is another interesting situation (which we will not discuss in
this paper) where finite size effects of
the type described in this paper can be observed. It is the cylindric
geometry described in ref.s~\cite{bbv,hp} in which two spatial directions
are chosen to be approximately of the same size as the inverse
deconfinement temperature.

\vskip 0.3cm
This paper is organized as follows:
in sect.2 we give a short discussion of the flux tube model,
in sect.3 we describe the details of the Monte Carlo simulation and
in sect.4 we  compare theoretical predictions and numerical results.

\section{ The fluctuating flux tube model}

The flux tube model is based on the idea that in the confined phase of
a gauge
theory the quark-antiquark pair should be joined together by a thin,
fluctuating flux tube~\cite{conj}. The simplest version of the model
(which should be a good description when heavy quark-antiquark
pairs are studied at large interquark separation $R$) assumes
that the chromo-electric flux is confined inside a tube of small but
nonzero thickness $L_c$, that $L_c$ is constant
along the tube (neglecting boundary effects near the quarks)
and independent of the interquark distance.

An immediate consequence of
this picture is linear confinement: the interquark potential $V(R)$
rises linearly according to the law $V(R)=\sigma R$.

\vskip .3 cm
A second, important consequence is that the string tension $\sigma$ and
the effective cross-section of the flux tube $A_t$ are related by the law
\eq
\sigma=\frac{c_t}{A_t}~~~~,
\label{cs}
\en
where $A_t$ is defined as
\eq
A_t\equiv\frac{\left(\int dA ~E \right)^2}{\int dA ~E^2 }~~~~.
\en
In some cases the constant $c_t$ and, in particular, its dependence
on the quark representation, can also be determined
(see ref.~\cite{tw} for a discussion on this point and for further
references).
If the chromoelectric flux $E$ is constant across the tube, then
the simple dimensional  relation $A_t~\propto ~L_c^{(d-2)}$ holds,
where $(d-2)$ is the number of transverse dimensions.

Eq.~(\ref{cs}) can be tested, for instance, by measuring the string
tension on asymmetric lattices, with one space direction, say $L_s$,
smaller than the others.
If $L_s>L_c$ no effect on the string tension is expected.
On the contrary, when $L_s<L_c$ the flux tube is squeezed and the string
tension increases. In particular, if we
 assume  that the flux density is
uniform inside the flux tube, then
eq.~(\ref{cs}) suggests that the
rising of  the string tension is linear and obeys the law

\eq
\sigma(L_s)=\frac{L_c}{L_s}\sigma(\infty)
\label{siglt}
\en
where $\sigma(\infty)$ denotes the string tension in the uncompressed
situation, namely for $L_s>>L_c$ (in the following we will denote
$\sigma(\infty)$ with $\sigma$ for brevity).

This is the simplest possible
assumption on the behaviour of the chromoelectric flux $E$
across the tube and as a consequence
the linear behaviour of eq.(\ref{siglt}) is not at all mandatory
and will have in general to be corrected when regions deep inside
the flux tube are probed. Nevertheless, it has
has been neatly observed recently in the case of
the (2+1) dimensional SU(2) model~\cite{t1}. The interesting aspect
of such a linear behaviour implied by this uniform flux distribution is
that it allows a rather precise
determination of the flux tube thickness $L_c$, which turns out to be in
good agreement with other independent evaluations. For instance,
in the above mentioned case of
 the (2+1) dimensional SU(2) gauge model, the value quoted by
Teper~\cite{t1}  agrees with the corresponding value obtained by
Trottier and Woloshyn~\cite{tw} with a different method.

\vskip .3 cm
The fact that the flux tube has a finite non-zero thickness must be
carefully taken into account to avoid
 systematic errors in the evaluation of the string
tension from numerical simulations. It is in fact clear from the above
discussion that only Wilson loops of size greater than $L_c$ must be
used to extract the string tension in order to avoid the appearance of
unphysical effects, in particular an artificial enhancement of the
string tension.
This is particularly important since small size loops, due to
their small relative errors, dominate  the fits and can strongly
influence the determination of $\sigma$. An important cross-check of
these systematic errors is given by looking at the variation of
$\chi^2/d.o.f.$ (from now on denoted by $\chi_r^2$~)
as a function of the minimal size $L_{min}$ of the Wilson loops
considered in the fits. Only when $L_{min}$  is of the order of $L_c$
good $\chi_r^2$ are obtained and  stable values of the string tension
are found .
Notice also that these systematic deviations from the area law at small
distances cannot be considered as lattice artifacts, because of their
good scaling behaviour \cite{cfgpv}. They could in principle  be used
as an independent way to estimate $L_c$.

Let us finally remark that the enhancement of the string tension at
short distances is a general phenomenon: it can
be precisely seen also in the case of 4d SU(2) and SU(3) gauge theories
(see for instance fig.2 of ref.~\cite{cfg}) and also when Polyakov line
correlations  are studied (see for instance
fig.1 and fig.4 of ref.~\cite{cfgv}).

\vskip .3 cm
A third, important, consequence emerges if one tries to take into
account the quantum fluctuations of the flux tube.
  Below the roughening transition the Wilson loop follows
the area law, which is the lattice counterpart of the ``classical''
linearly rising potential, but the continuum limit, hence the connection
with the physical flux tube, can be achieved only after the roughening
transition, where the quantum fluctuations of the surface bordered by
the loop play a crucial role.
It is by now generally accepted that these  quantum fluctuations can be
effectively described  by a  massless two-dimensional free field theory.
This idea traces back to the seminal work by L\"uscher, Symanzik
and Weisz~\cite{lsw} and has been discussed from then in several
papers both in
the context of lattice gauge theories (see for instance
ref.~\cite{lat92} and references therein) and in the dual context of
interfaces in 3d spin systems (see for instance
ref.~\cite{interface} and references therein).
Let us briefly review this approach and fix notations and
conventions.

\subsection{Quantum fluctuations of the flux tube}

The starting point  is the assumption that
the fluctuations of the surface bordered by the Wilson loop are
described by an effective Hamiltonian proportional to the change they
produce in  the area of the surface itself
\eq
H_{eff}=\sigma\int_0^{L_1} dx_1
\int_0^{L_2}dx_2~ \left[\sqrt{1+\left(\frac{\partial h}{\partial x_1}\right)^2
+\left(\frac{\partial h}{\partial x_2}\right)^2}-1\right]~~,
\label{cw1}
\en
where the field $h(x_1,x_2)$ describes the surface displacement from
the equilibrium position as a function of the longitudinal coordinates
$x_1$ and $x_2$, $L_1~(L_2)$ is the
size of the Wilson loop in the $x_1~(x_2)$ direction and $\sigma$ is the
string tension \footnote{Note, as a side remark, that
eq.~(\ref{cw1}) coincides
with the Nambu-Goto string action in a special frame where only a subset
of surface configurations are allowed. However the results of this
section are also valid in a more general frame where all the possible
configurations are taken into account.}.

The contribution to the Wilson loop expectation value  due to surface
fluctuations is then given by

\eq
Z_{eff}=~ tr~e^{-H_{eff}}~~~~.
\label{free}
\en
The Hamiltonian of eq.~(\ref{cw1}) is  too difficult to be handled exactly.
However a crucial observation is  that this theory can be expanded in
the adimensional parameter $({\sigma L_1L_2})^{-1}$ and the leading order
term is the gaussian model,
which will be a good approximation when large enough Wilson loops are
studied. Then we replace eq.~(\ref{cw1}) with the $\sigma L_1L_2
\to\infty$ limit $H \to H_G$

\eq
H_G=\frac{\sigma}{2}
\int_0^{L_1} dx_1 \int_0^{L_2} dx_2 ~ \left[\left(\frac{\partial h}
{\partial x_1}\right)^2
+\left(\frac{\partial h}{\partial x_2}\right)^2\right]~~~~.
\label{cw2}
\en
Within this approximation, the integration over $h$ implied in eq.~
(\ref{free}) can be done exactly. The integral is divergent
but can be regularized using, for instance, a suitable generalization of
the Riemann $\zeta$-function regularization
(see $e.g.$~\cite{id}).
The result  depends only on the geometrical properties of the boundary.
In particular, for a Wilson loop with fixed (``Dirichlet'') boundary
conditions along the loop, the gaussian contribution turns out to be
\eq
Z_G(L_1,L_2)=\frac{c}{\sqrt{\eta(\tau){\sqrt{L_2}}}}\hskip0.5cm
,\hskip0.5cm \tau=i{L_1\over L_2}~~~,
\label{bos}
\en
\noindent
where $\eta$ denotes the Dedekind eta function
\eq
\eta(\tau)=q^{1\over24}\prod_{n=1}^\infty(1-q^n)\hskip0.5cm
,\hskip0.5cmq=e^{2\pi i\tau}~~~,\label{eta}
\en
$c$ is an undetermined constant and we have assumed $L_1\geq L_2$,
without loss of generality.

Thus, taking into account the area term $\sigma L_1L_2$
(coming from the classical or zeroth order contribution of the saddle
point) and the undetermined  perimeter contribution $p(L_1+L_2)$ (which
depends on the lattice regularization), the Wilson loop expectation
value $W(L_1,L_2)$ may be written in the form
\eq
W(L_1,L_2)={\rm e}^{-\sigma L_1L_2+p(L_1+L_2)}Z_G~~ ,
\en
where the  contribution $Z_G$ of the quantum fluctuations
can be expanded as follows

\eq
Z_G(L_1,L_2)= \frac{c}{\sqrt[4]{L_2}}
 q^{-{\frac{1}{48}}} \sqrt{1+q+2q^2+3q^3+5q^4+\dots}
\label{zcw}
\en
with $q=exp(-2\pi\frac{L_1}{L_2})$. Eq.~(\ref{zcw}) is symmetric under
the exchange $L_1\leftrightarrow L_2$, as one can check by applying the
modular transformation $\tau \to -\tau^{-1}$ at eq.~(\ref{bos})
(see $e.g.$~\cite{id}).

As a consequence, a simple area-perimeter-constant
law cannot fit the Wilson loop expectation values, and this
shows up in  high $\chi_r^2$'s and unacceptable confidence
levels\footnote{Notice, for completeness, that in the case of  gauge
theories with continuous gauge groups,
perturbative contributions (for instance one gluon exchange terms)
are also expected  for small interquark separations.
However these are completely independent from the above discussed
infrared contributions and should be separately taken into account
in fitting Wilson loop expectation values~\cite{cfg}.
Notice also that the above discussion requires non-trivial modifications
if ``fuzzy'' or ``blocked'' versions of the Wilson loop are studied.}.
On the contrary, if the contribution of the
fluctuations  is taken into account,
 impressive reductions of the $\chi_r^2$ are found, and eventually
acceptable confidence levels are reached (see for instance~\cite{cfgpv}).
This is up to now one of the strongest evidences for the correctness of
this  description of the fluctuations of the  flux tube.

\vskip .3 cm
One of the consequences of the above description in terms of a free
bosonic field is that the mean width of the flux tube is expected
to grow logarithmically as a function of the interquark distance
\cite{lsw}, while there are reasons to believe that  such a width
should be constant (see for instance Ref.\cite{fstring}).
Actually there is  a slight modification of this theory which  accounts
for this fixed thickness, based on the observation that the free boson
can be seen as a limit of a one-parameter family of conformal field
theories, where this parameter can be simply related to the width $L_0$
of the flux tube.

In Ref.\cite{fstring} it has been argued that the value of
this parameter can be determined by matching the boundary conditions of
the conformal theory with those of the underlying gauge field theory,
combined with
the obvious assumption that the flux tube cannot self-overlap freely.
The latter assumption can be formulated more precisely in the Ising
gauge theory by saying that the flux tube must sweep in its time
evolution self-avoiding surfaces (see Ref.~\cite{lat92} for a
discussion of this point).

In this way one finds \cite{fstring}   $L_0=\sqrt{\pi/4\sigma}~$ for a
flux tube {\it at zero physical temperature}.
In the following we shall not use this result, because we want to
evaluate such a width by applying a completely independent argument
to a different physical situation (the flux tube at the critical
temperature $T^c$). We anticipate that the resulting value of $L_0$
at $T^c$ turns out to be very near the one quoted above.

Another consequence of this  mentioned modification of the free boson
theory is that the  contribution due to the quantum fluctuations of
the flux tube has now a different form:

\eq
Z_{SA}(L_1,L_2)=\tilde c
 q^{-{\frac{1}{192}}}
     \sqrt[4]{1+q^{\frac{1}{2}}+q^{\frac{3}{2}}+q^{2}+\dots}
\label{zsa}
\en
where $\tilde c$ is an undetermined constant and
$q=exp(-2\pi\frac{L_1}{L_2})$.

Notice however that, while
this modification has important consequences in the low temperature,
 confining regime, it is
actually irrelevant in the high temperature phase we are discussing in
this paper. Hence we will only refer to this choice when
comparing our results with data taken in the low temperature
phase.

\subsection{ Finite Temperature}

The fact that the flux tube has a non-zero thickness
 becomes particularly relevant when the Wilson loop is studied at a
finite temperature $T$, because this
introduces into the game a new scale $i.e.$ the length of the lattice
in the time direction $L_t=1/T$. For high enough temperatures,
this length eventually becomes comparable with the finite thickness of
 the flux tube and the  free field picture described above breaks down.

Notice however that, since we are studying Wilson loops orthogonal
to the time direction, this ``temperature'' interpretation is not
mandatory.
A space-like Wilson loop at high temperature is completely equivalent
to an ordinary Wilson loop in a  zero  temperature
environment, with the lattice size in one of the remaining space-like
directions smaller than the other. This is exactly the situation
described at the beginning of this section and studied by Teper
in~\cite{t1}: all the results listed there still hold with the
simple exchange of $L_s$ with $L_t$.
Exactly as before, choosing higher temperatures,
namely looking at smaller and smaller values of
$L_t$, we are actually probing the interior of the flux tube.

In finite temperature LGT's the
interquark potential can be extracted only by looking at the
correlations of Polyakov loops. The surface bordered by two Polyakov
loops undergoes a roughening transition exactly as in the Wilson loop
case and the same arguments and techniques described above for the
quantum fluctuations of
Wilson loops apply also in this case~\cite{ap,olesen}.
Using these results
Olesen was able to predict the ratio between the deconfinement
temperature and the square root of the zero-temperature
(namely $L_t=\infty$) string tension
$\sigma$~\cite{olesen}

\eq
\frac{T^c}{\sqrt{\sigma}}=\sqrt{\frac{3}{\pi(d-2)}}
\label{ol}
\en
where $(d-2)$ is the number of transverse dimensions of the flux tube.
This prediction turns out to be in good agreement with the results
obtained with Monte Carlo simulations in  (3+1) dimensions, both for the
 SU(2) and SU(3) gauge theories. The comparison with the numerical data
for models in (2+1) dimensions is less good but the disagreement is
anyway contained within $20\%$. For a review of these data see
 tab.III and ref.~\cite{lat92}.

\subsection{Vanishing of the quantum fluctuations}

Let us make at this point our main assumption.
We assume
that the
{\sl quantum corrections $Z_G$ should disappear} when the flux tube
fills the whole lattice, namely when $T>T^c$ or
$L_s<L_c$, depending on the geometry we are interested in.
Intuitively this is equivalent to assume some sort
of self-avoiding behaviour of
the flux tube, since in this case , when the flux tube fills the whole
lattice  there is no more space left
for it to fluctuate. We will show below that
this assumption is satisfied in the case of the (2+1) $Z_2$ model.
This vanishing of quantum fluctuations can be described
in a more rigorous way by
noticing that the compactification in one lattice direction (say, $L_s$)
 naturally induces a
compactification of the field $h(x_1,x_2)$ on a circle of radius
$R=\frac{L_s}{2\pi}$. The quantum field theory of
a two-dimensional bosonic field compactified on a circle is by now
rather well understood. In particular
the spectrum of states and the partition function are exactly known (at
least for the so called ``rational models'' for which $R^2$ is a
rational number). It is thus possible to follow the behaviour of the
quantum corrections as a function of $R$: $Z_G=Z_G(R)$. What is
interesting  is that there are precisely four
values of $R$ (but only two of them are independent, the others being
related by duality) for which the contribution of such quantum
corrections vanishes (they
correspond to the so called ``topological field theories'', see
ref.~\cite{lat92} for notations and bibliography on this subject).
According to ref.~\cite{lat92,tc},
we can then predict that the value (let us call it $L_0=2\pi R_0$)
of the lattice size which corresponds to a zero-contribution point must
be related to the (zero temperature) string tension as follows
\eq
L_0=\sqrt{\frac{\pi}{3\sigma}} ~~~.
\label{n=2}
\en

Following our assumption we can thus say
that {\sl  moving toward higher temperatures  corresponds in the
flux tube effective model to a flow toward one of these
zero-contribution points}  and that {\sl $L_0$ must coincide with
$L_c$}, the
flux tube thickness.  By comparing eq.s (\ref{n=2}) and (\ref{ol}) we then
 see that in (2+1) dimension the flux tube thickness $L_c$
 coincides  with the inverse deconfinement temperature.

Eq.(\ref{n=2}), and the fact that $L_c=1/T^c$ are in rather good
agreement (within at most a $20\%$ of deviation depending on the model)
with the result of the simulations  of
the (2+1) dimensional SU(2) model~\cite{t1,t2} and with our simulations
of the (2+1) $Z_2$ gauge model (see tab.III). According to the
subsection [d] of the introduction we suggest that this similarity
should hold for any gauge theory in (2+1) dimensions.

 This similarity allows us to identify the
deconfined regime with the compressed flux tube regime
and to interpret {\em the increasing of the space-like string tension
in the deconfined phase as a signature of the compression of the flux
tube}.

\section{ (2+1) $Z_2$ gauge Monte Carlo simulation}

We compared our predictions with a set of high statistics simulations
of the  $Z_2$ gauge model in (2+1) dimensions. Let us briefly list some
of the reasons for this choice (a more detailed discussion can be found
for instance in ref.s~\cite{cfgpv,cfgv}).

\begin{description}
\item{a]}
Due to the fact that this model
is the dual of the 3d Ising model, very precise
values of all bulk quantities (critical couplings and indices)
are known. Moreover,  in the region that we
study an excellent agreement with the scaling laws has been found.

Choosing the standard normalization for the action
\eq
S=-\beta \sum_p U_p~~,
\en
where $U_p$ denotes the product of $Z_2$ elements associated to the
links
belonging to the plaquette $p$, we have a roughening
transition at $\beta_R\simeq0.4964$, a (zero temperature) deconfinement
transition at  $\beta_c\simeq0.7614$ and a critical index for
the correlation function $\nu\simeq 0.63$.
Precise information on the finite temperature
deconfinement transition can be found, for instance, in~\cite{wz}.
The scaling region of this transition starts approximatively at $L_t=4$.
Data obtained with the Monte Carlo renormalization group approach for
$L_t=4$ and $L_t=8$ give: $\beta_c(L_t=4)=0.7315(5)$ and
$\beta_c(L_t=8)=0.7516(5)$~\cite{wz}
(where the notation $\beta_c(L_t)$ denotes
the value of $\beta$ at which deconfinement occurs in a lattice with
temporal size $L_t$). Defining the critical temperature
$T^c$ as
\eq
\frac{1}{L_t(\beta)}~=~T^c_0~(\beta_c-\beta(L_t))^{\nu}~~~,
\label{tc}
\en
we have $T^c_0=2.3~(1)$.

\item{b]}
The model is simple enough to allow high statistics simulations, but
shares the same infrared behaviour (in the $\beta<\beta_c$ region) with
non abelian lattice gauge theories.
In particular, near the finite temperature deconfinement point, the
Svetitsky-Yaffe conjecture tells us that $Z_2$ and SU(2) models should
be in the same universality class.

\item{c]}
Recently the interface
behaviour in 3d statistical models has attracted a lot of interest.
 In d=3 the physics of interfaces is
exactly equivalent, through duality, to  the physics of Wilson loops
and, in the particular case of the Ising model, not only theoretical
tools but also
numerical results can be borrowed from one context to the other (see
ref.~\cite{interface} for a discussion of this problem). In particular,
high precision estimates of the string tension $\sigma$
(equivalent to the interface tension) exist. In the scaling region
the string tension behaves as
\eq
\sigma(\beta)~=~\sigma_0~(\beta_c-\beta)^{2\nu}~.
\label{scal}
\en
It turns out that this law is very well fitted by the existing
data~\cite{interface} with $\sigma_0=3.70~(4)$.
By inserting the value for the critical temperature obtained from
eq.~(\ref{tc}), we can construct the adimensional ratio:
${T^c_0}/{\sqrt{\sigma_0}}=1.19~(6)$.

\end{description}

\subsection{The simulation}

The simulation was performed on a $48^2\times L_t $ lattice
($L_t=$ lattice spacing in the time direction) with periodic boundary
conditions in all space-time directions.
A standard heat-bath algorithm was used to
update links. Four values of $\beta$, with various choices of $L_t$,
were studied. They are listed in tab.I~.

%     TAB 1
\vskip0.3cm
$$\vbox {\offinterlineskip
\halign  { \strut#& \vrule# \tabskip=.5cm plus1cm
& \hfil#\hfil
& \vrule# & \hfil# \hfil &
& \vrule# & \hfil# \hfil &\vrule# \tabskip=0pt \cr \noalign {\hrule}
&& $\beta$ && $L_t$  && $L_c(\beta)$ & \cr \noalign {\hrule}
&& $0.7420$ && $13$  && $5.2~(2)$ & \cr \noalign {\hrule}
&& $0.7500$ && $10,12$  && $7.3~(3)$ & \cr \noalign {\hrule}
&& $0.7525$ && $2-8$  && $8.5~(4)$ & \cr \noalign {\hrule}
&& $0.7585$ && $2-16$  && $17.2~(7)$ & \cr \noalign {\hrule}
}}$$
\begin{center}
{\bf Tab.I.}{\it ~Set of measured data. In the third column,
the corresponding inverse critical temperatures (in unities of the
lattice spacing) are presented.}
\end {center}
\vskip0.3cm

The reason for these choices is the following.
The data at $\beta=0.7585$
have been measured to make quantitative estimates of the temperature
dependence of $\sigma(L_t)$. Since in this case the
 finite temperature deconfinement point is at $L_t\sim 17$,  a wide
range of values of $L_t$ can be used to check the predicted behaviour of
the string tension.
The data at $\beta=0.7525$ have been taken to check the scaling behaviour
and the overall reliability of our results.
In both cases
all deconfined values of $L_t$ were studied,
 namely $2\leq L_t\leq 8$ for
$\beta=0.7525$ and  $2\leq L_t\leq 16$ for
$\beta=0.7585$.

Finally, to make a comparison with the above data, we
studied three samples in the confining phase
at intermediate finite temperatures (from $T/T^c\sim 0.4$ to
$T/T^c\sim 0.73$): these have been chosen in order
to have low enough temperature to guarantee an uncompressed
behaviour of the flux tube (in particular it can fluctuate freely).
The values of $\beta$ ($\beta=0.742$ and  $\beta=0.750$)
have been tuned so as to have values of the string tension
comparable with those of the compressed case.
In particular, $\sigma(\beta=0.742)\sim
\sigma(\beta=0.7525,L_t=4)\sim\sigma(\beta=0.7575,L_t=4)$ and
  $\sigma(\beta=0.7575,L_t=6) <\sigma(\beta=0.750)\sim
\sigma(\beta=0.7525,L_t=6) < \sigma(\beta=0.7575,L_t=5)$,
(see tab.II$a$,$b$,$c$).

All Wilson loops orthogonal to the time direction $W(L_1,L_2)$
in the range $2\leq L_1, L_2\leq 20$ were measured.
Also this choice  requires some explanation. It is well known
that improved versions of the Wilson loop operator~\cite{ppr}
give more precise results, but notice that only in the case of ordinary
Wilson loops the
gaussian determinants described in the previous section can be evaluated
exactly.
Since our goal is to show the fate of the
quantum fluctuations of the flux tube across the deconfinement
transition, more than to have precise evaluations of, say, $L_c$,
we decided to concentrate on ordinary, not improved, Wilson loop
expectation values.

\subsection{ The cross-correlation problem}

The major problems one has to face for these values of $\beta$
 are the  critical slowing down and the huge
cross-correlations among Wilson loops. The first problem was kept under
control separating each measurement on the lattice with 32 sweeps. The
second was taken into account, as usual, by  weighting the data in the
fitting procedure with the inverted cross-correlation matrix . In the
most severe cases, namely those near the critical point ($L_t=6-8$ for
$\beta=0.7525$, and $L_t=13-16$ for $\beta=0.7585$)
  a scattering procedure was also implemented, measuring
in each iteration only one single Wilson loop of fixed size and
scattering the measure of the others in the Monte Carlo
time\footnote{In these cases the cross-correlation matrix was
so flat, within the statistical errors, that the reliability of the
whole inversion procedure and cross-correlated fit calculation, without
a proper scattering procedure, was rather doubtful.}.
Let us stress that all these steps were absolutely crucial  to obtain
reliable confidence levels in the fits.
Moreover in these last cases further runs with lattice size $96^2\times
L_t$ were performed, measuring loops up to $40 \times 40$, to check the
reliability of our results. Notice finally, as a side remark,
that improved Wilson loop estimators experience even worse
cross-correlations.

All the quoted errors were obtained with a standard jacknife procedure.

\section{ Results and conclusions}

For each value of $\beta$ and $L_t$ we fitted the Wilson loop
expectation values with a pure area law
\eq
W(L_1,L_2)=exp\{-\sigma L_1L_2~+~p(L_1+L_2)~+~c\}
\en
(we shall call this choice ``type 1'' fit in the following)
and with an area law corrected in order to take into account the
quantum fluctuations of the Wilson loop surface
\eq
W(L_1,L_2)=Z_G(L_1,L_2)~exp\{-\sigma L_1L_2~+~p(L_1+L_2)~+~c\}
\en
(``type 2'' fit in the following).
For the three samples in the confined phase we fitted the data also with
the modified version of the gaussian contribution  proposed in
eq.~(\ref{zsa})

\eq
W(L_1,L_2)=Z_{SA}(L_1,L_2)~exp\{-\sigma L_1L_2~+~p(L_1+L_2)~+~c\}
\en
(``type 3'' fit in the following).

We performed the fits setting a lower threshold $R_t$ in the
size of the Wilson loops. To be precise we constructed the following
subsets of our Wilson loops samples
\eq
S(R_t)=\{W(L_1,L_2),~ L_1\geq L_2\geq R_t\}
\en

\noindent
choosing $R_t=L_t$ in the deconfined phase.
In the confined region we chose instead $R_t=L_c(\beta)$ (namely
the flux tube thickness, extrapolated through scaling at those values of
$\beta$).
This means $R_t=4$ for $\beta=0.742$ and $R_t=6$ for $\beta=0.750$.
Notice that, as a consequence of this cutoff procedure, severe problems
of precision  exist   near the critical temperature.
The results of the fits are collected in tab.II$a$,$b$,$c$ and fig.1,2.

Let us make some comments on these data.
\begin{description}

\item{\sl $\chi_r^2$ behaviour}

In the deconfined phase type 1 fits have in general better confidence
levels than type 2 fits (see tab.II$a$ and tab.II$b$).
This is quite evident in the high temperature regime $T>2T^c$, while
near the critical point the two $\chi_r^2$'s almost  coincide, indicating
that the quantum fluctuations corrections in these region are smaller
than the precision of our data and  could not be   detected in any case
even if they are present.
This behaviour is even more
distinct if compared with analogous fits for the two samples of data
at low temperature, in the confined region (see tab.II$c$), where
definitely better confidence levels are obtained with fits of type 2
and 3.
This is a first signature of the vanishing of quantum fluctuations.

\item{$T^c<T<2T^c$}

Let us concentrate on the data at $\beta=0.7585$ (tab.II$a$).
We  fitted the values of $\sigma(L_t)$ in the range $9\leq L_t\leq 16$
according to the linear
law of eq.~(\ref{siglt}). As it can be seen in fig.1, the linear
law is in good agreement with the data, and in fact the fit shows
a good confidence level:
 C.L.=98\%. As a result we obtain $L_c=18.5~(9)$, which implies:
$L_c=1.07(5)(5){T^{-1}_c}$ and $L_c\sqrt{\sigma_0}=0.90(5)(2)$,
where the second error in these relations is due respectively to the
uncertainty in the critical temperature and the string tension.

\newpage

%     TAB 2a
$$\vbox {\offinterlineskip
\halign  { \strut#& \vrule# \tabskip=.5cm plus1cm
& \hfil#\hfil & \vrule# & \hfil# \hfil
& \vrule# & \hfil# \hfil
& \vrule# & \hfil# \hfil
& \vrule# & \hfil# \hfil &
& \vrule# & \hfil# \hfil &\vrule# \tabskip=0pt \cr \noalign {\hrule}
&& $L_t$ && $T/T^c$  && $\sigma$  && $\sigma_0$
&& $\chi^2_1$ && $\chi^2_2$ & \cr \noalign {\hrule}
&& $2$ && $8.6~(4)$ && $0.120~~(3)$ &&
$189.~(5)$&&$1.3$ && $7.1$ & \cr \noalign {\hrule}
&& $3$ && $5.7~(3)$ && $0.0480~~(2)$ &&
$75.6~(3)$ &&$0.71$&& $8.1$ & \cr \noalign {\hrule}
&& $4$ && $4.3~(2)$ && $0.0244~(3)$ &&
$38.8~(5)$ &&$0.85$&& $5.1$ & \cr \noalign {\hrule}
&& $5$ && $3.45~(17)$ && $0.0150~(4)$ &&
$23.6~(6)$ &&$1.2$&& $1.8$ & \cr \noalign {\hrule}
&& $6$ && $2.87~(14)$ && $0.0100~(3)$ &&
$15.8~(5)$ &&$0.91$&& $1.8$ & \cr \noalign {\hrule}
&& $7$ && $2.46~(12)$ && $0.0083~(5)$ &&
$13.1~(8)$ &&$1.3$&& $1.3$ & \cr \noalign {\hrule}
&& $8$ && $2.15~(11)$ && $0.0068~(4)$ &&
$10.7~(6)$ &&$0.78$&& $1.3$ & \cr \noalign {\hrule}
&& $9$ && $1.92~(10)$ && $0.0049~(4)$ &&
$7.7~(6)$ &&$0.90$&& $1.08$ & \cr \noalign {\hrule}
&& $10$ && $1.73~(9)$ && $0.0044~(4)$ &&
$6.9~(6)$ &&$1.1$&& $1.4$ & \cr \noalign {\hrule}
&& $11$ && $1.57~(8)$ && $0.0037~(4)$ &&
$5.8~(6)$ &&$0.72$&& $0.80$ & \cr \noalign {\hrule}
&& $12$ && $1.44~(7)$ && $0.0038~(4)$ &&
$6.0~(6)$ &&$1.01$&& $1.06$ & \cr \noalign {\hrule}
&& $13$ && $1.33~(7)$ && $0.0033~(5)$ &&
$5.2~(8)$ &&$0.90$&& $0.90$ & \cr \noalign {\hrule}
&& $14$ && $1.23~(6)$ && $0.0033~(5)$ &&
$5.2~(8)$ &&$0.52$&& $0.74$ & \cr \noalign {\hrule}
&& $15$ && $1.15~(6)$ && $0.0025~(8)$ &&
$3.9~(1.3)$ &&$0.46$&& $0.48$ & \cr \noalign {\hrule}
&& $16$ && $1.08~(5)$ && $0.0022~(9)$ &&
$3.5~(1.4)$ &&$0.85$&& $0.85$ & \cr \noalign {\hrule}
}}$$
\vskip-0.2cm
\centerline {\bf a}

%     TAB 2b
$$\vbox {\offinterlineskip
\halign  { \strut#& \vrule# \tabskip=.5cm plus1cm
& \hfil#\hfil & \vrule# & \hfil# \hfil
& \vrule# & \hfil# \hfil
& \vrule# & \hfil# \hfil
& \vrule# & \hfil# \hfil &
& \vrule# & \hfil# \hfil &\vrule# \tabskip=0pt \cr \noalign {\hrule}
&& $L_t$ && $T/T^c$  && $\sigma$  && $\sigma_0$
&& $\chi^2_1$ && $\chi^2_2$ & \cr \noalign {\hrule}
&& $2$ && $4.2~(2)$ && $0.128~~(2)$ &&
$49.1~(8)$&&$0.79$ && $2.73$ & \cr \noalign {\hrule}
&& $3$ && $2.84~(14)$ && $0.0520~~(4)$ &&
$19.9~(2)$ &&$0.81$&& $1.68$ & \cr \noalign {\hrule}
&& $4$ && $2.13~(10)$ && $0.0270~(10)$ &&
$10.0~(4)$ &&$1.08$&& $1.44$ & \cr \noalign {\hrule}
&& $5$ && $1.70~(8)$ && $0.0172~(5)$ &&
$6.6~(2)$ &&$0.96$&& $1.00$ & \cr \noalign {\hrule}
&& $6$ && $1.42~(7)$ && $0.0137~(7)$ &&
$5.3~(3)$ &&$0.70$&& $0.81$ & \cr \noalign {\hrule}
&& $7$ && $1.22~(6)$ && $0.0112~(7)$ &&
$4.3~(3)$ &&$0.80$&& $0.80$ & \cr \noalign {\hrule}
&& $8$ && $1.06~(5)$ && $0.0085~(9)$ &&
$3.3~(3)$ &&$0.70$&& $0.70$ & \cr \noalign {\hrule}
}}$$
\vskip-0.2cm
\centerline {\bf b}

%     TAB 2c
$$
\hskip-2cm
\vbox {\offinterlineskip
\halign  { \strut#& \vrule# \tabskip=.5cm
& \hfil#\hfil & \vrule# & \hfil# \hfil
& \vrule# & \hfil# \hfil
& \vrule# & \hfil# \hfil
& \vrule# & \hfil# \hfil
& \vrule# & \hfil# \hfil
& \vrule# & \hfil# \hfil &
& \vrule# & \hfil# \hfil &\vrule# \tabskip=0pt \cr \noalign {\hrule}
&& $\beta$ && $L_t$ && $T/T^c$  && $\sigma$  && $\sigma_0$
&& $\chi^2_1$ && $\chi^2_2$&& $\chi^2_3$ & \cr \noalign {\hrule}
&& 0.742&& $13$ && $0.40~(2)$ && $0.027~~(2)$ &&
$3.88~(20)$&&$1.7$ && $0.85$&& $0.80$ & \cr \noalign {\hrule}
&& 0.750&& $12$ && $0.61~(3)$ && $0.0132~~(10)$ &&
$3.71~(29)$&&$1.4$ && $1.1$&& $1.02$ & \cr \noalign {\hrule}
&& 0.750&& $10$ && $0.73~(3)$ && $0.0127~~(11)$ &&
$3.56~(32)$&&$1.4$ && $1.1$&& $0.89$ & \cr \noalign {\hrule}
}}$$
\vskip-0.2cm
\centerline {\bf c}

\begin{center}
{\bf Tab.II.~(a,b,c)}
{\it~Values of the string tension extracted from Wilson loops at
$\beta=0.7585$ (a); $\beta=0.7525$ (b);
$\beta=0.7420,~0.7500$ (c).
In the first two columns the lattice size in the time direction and the
corresponding temperature (in units of the critical
temperature) are reported. The last columns contain the reduced $\chi^2$
of the fits of type 1 and 2 (in tab.IIc also fits of type 3 are
considered). The third  column contains the string
tension $\sigma(L_t)$,  extracted from  fits of type 1 in the cases
a,b and fits of type 3 in the case c. In the fourth column are reported
the corresponding scaling values $\sigma_0(L_t)$, according to
eq.~(\ref{scal}).}
\end {center}
\newpage

A slightly lower confidence level (35\%) is obtained if the same fit is
performed on the $\beta=0.7525$ data (tab.II$b$), with $L_c=8.7(5)$,
which corresponds to $L_c=1.02(5)(5){T^{-1}_c}$ and
$L_c\sqrt{\sigma_0}=0.86(5)(2)$.

These results are summarized in tab.III, where they are also compared
with the corresponding values for the SU(2) model~\cite{t1,t2} and with
our predictions eq.s(\ref{ol}) and (\ref{n=2}).

%     TAB 3
\vskip0.3cm
$$\vbox {\offinterlineskip
\halign  { \strut#& \vrule# \tabskip=.5cm plus1cm
& \hfil#\hfil
& \vrule# & \hfil# \hfil
& \vrule# & \hfil# \hfil &
& \vrule# & \hfil# \hfil &\vrule# \tabskip=0pt \cr \noalign {\hrule}
&& $model$ && $L_c\sqrt{\sigma}$  && ${T^c}/{\sqrt{\sigma}}$
&& $L_cT^c$& \cr \noalign {\hrule}
&& $SU(2)$ && $\sim 1.09$  && $1.12(1)$
&& $\sim 1.22$ & \cr \noalign {\hrule}
&& $Z_2$ && $0.90(7)$ && $1.19(6)$  && $1.07(10)$ & \cr \noalign {\hrule}
&& $FTM$ && $1.024$  && $0.97$&&$1$ & \cr \noalign {\hrule}
}}$$
\begin{center}
{\bf Tab.III.}{\it ~Comparison between $Z_2$ and $SU(2)$ gauge theories
in (2+1) dimensions. In the last row are reported the flux tube model
(FTM) predictions eq.s(\ref{ol}) and (\ref{n=2}).}
\end {center}
\vskip0.3cm

\item{$T>2T^c$}

In this region the linear behaviour of eq.~(\ref{siglt}) is completely
lost. This indicates the appearance of some non-trivial structure deep
inside the flux tube. The data show a good scaling behaviour as a function
of $T$ and suggest  that the only remaining physical scale in this
regime is the temperature.

This can be seen by looking at fig.2.
Since the errors in this region are very small, the $\chi_r^2$ becomes a
very efficient tool to select among various possible behaviours. Indeed,
as can be seen in fig.2, both the data at $\beta=0.7525$ and those at
$\beta=0.7585$ have a very precise power law behaviour
\eq
\sigma(L_t)=a~L_t^{\alpha}
\label{hight}
\en

and  the two values of $\alpha$ and $a$ at the two
$\beta$'s are almost compatible within the errors.

The output of the fits is

$\beta=0.7525$, range: $4\geq L_t\geq 2$,   C.L.$=66$\%, $a=0.58(2)$,
$\alpha=-2.19~(4)~~,$

$\beta=0.7585$, range: $6\geq L_t \geq 2 $,   C.L.=$75$\%, $a=0.59(2)$,
$\alpha=-2.28~(4)$~~.

\end{description}

\subsection{ Vanishing of quantum fluctuations}

In order to have an independent check of the vanishing of quantum
fluctuations in the deconfined phase we constructed the ratios
\eq
C(L,n)=\frac{W(L,L)}{W(L+n,L-n)}~~~~.
\label{r1}
\en
If the Wilson loops are described by a pure area law then
$C(L,n)$ does not depend on $L$
\eq
C(L,n)={\rm e}^{-\sigma n^2}
\en
while, if a contribution coming from the flux tube fluctuations is
present, a decreasing function of $L$ is expected
\eq
C(L,n)=\frac{Z_G(L,L)}{Z_G(L+n,L-n)}{\rm e}^{-\sigma n^2}
\label{sg}
\en
or, following eq.~(\ref{zsa}),
\eq
C(L,n)=\frac{Z_{SA}(L,L)}{Z_{SA}(L+n,L-n)}{\rm e}^{-\sigma n^2}~~~~.
\label{ssa}
\en

Note that, being only two Wilson loops involved in these ratios,
one has much smaller errors than, for instance,  in the case of  Creutz
ratios. We decided to compare samples of data with (almost) the same
string tension and the same expected flux tube thickness
in order to minimize systematic errors (due, for instance, to
different lower thresholds in fitting the data). Hence we compared the
Wilson loop expectation values at $\beta=0.7585,~~L_t=4$ and
$\beta=0.7525,~~L_t=4$ with those at $\beta=0.7420,~~L_t=13$. It can be
neatly seen by looking at fig.3 that the two samples in the deconfined
phase (compressed flux tube)  have almost coinciding ratios and show no
contribution coming from flux tube fluctuations, while the ratios of the
sample  in the confining region lie in between the slopes predicted by
eq.~(\ref{sg}) and eq.~(\ref{ssa}). The same picture is confirmed by fig.4,
where the samples at $\beta=0.7585,~~L_t=5,6$ and $\beta=0.7525,~~L_t=6$
are compared with those at $\beta=0.7500,~~L_t=10,12$.
Notice that in this case the lower threshold is $R_t=6$, nevertheless
 also the data at lower values of $R$ have been included in the figure,
in order to show the enhancement of the string tension below the
threshold, as discussed in sect.2.
Note also that the proposal of
eq.~(\ref{zsa}) and eq.~(\ref{ssa}) for the flux tube fluctuations
seems to describe better
the data exactly in that region in which the self-avoiding constraint
plays a major role.

Thus we can conclude that the information extracted
from $C(L,n)$ is in complete agreement with
those coming from the $\chi_r^2$ of the fits and support the picture
of the vanishing of quantum flux tube fluctuations in the
high temperature phase.

\subsection{ Comparison with the SU(2) gauge theory}

A remarkable feature of our results is their similarity with those
obtained by Teper in the case of SU(2) in d=2+1 dimensions (see tab.III).
Indeed, this also is a consequence of the flux tube picture, which, at
least in a first order approximation, does not depend on the ultraviolet
details of the model (not even on the fact that the gauge group is
discrete or continuous, abelian or non-abelian), but only on its
infrared behaviour (hence essentially on the existence of a confining
phase beyond the roughening transition and on the space time
dimensions). Let us stress that the comparison made in tab.III is
only a partial check of this supposed universality, since both the SU(2)
and the Ising model have the same center.
It would be quite interesting to check it  also in other cases, and
in particular for the SU(3) model.

\vskip .5cm
{\bf Acknowledgements}
\vskip .2cm
One of us (S.V.) would like to thank the CBPF for the
kind and warm hospitality in Rio de Janeiro.
The work of S.V. in Rio de Janeiro has been supported by a
CNPq grant.

\vskip 2cm

\begin{center}
{\bf Figure Captions}
\end{center}

\vskip0.8cm

{\bf Fig.1.}
{\it~ The string tension in scaled units ($\sigma_0(L_t)$ in tab.II$a$
,$b$,$c$) is plotted as a function of
$T/T^c$. Data corresponding to tab.II$a$ (triangles), tab.II$b$
(squares),
tab.II$c$ (crosses),  are reported. The horizontal line is the
uncompressed value of the string tension $\sigma_0=3.70$. The other line
is the best fit to eq.~(\ref{siglt}) for the sample of
data at $\beta=0.7585$ in the range $9\leq L_t\leq 16$.}
\vskip0.6cm

{\bf Fig.2.}{\it~ The string tension ($\sigma(L_t)$ in tab.II$a$
,$b$) is plotted as a function of
$T$ ($1/L_t$ in tab.II$a$,$b$). Data corresponding to tab.II$a$
(triangles) and tab.II$b$ (squares),
  are reported.
Both $\sigma(L_t)$ and $T$ are in log. scale.  The
continuum line is the best fit to eq.~(\ref{hight}) for the sample of
data at $\beta=0.7585$ in the range $2\leq L_t\leq 6$ . }
\vskip 0.6cm

{\bf Fig.3.}
{\it~ The logarithm of the Wilson loop ratios of eq.~(\ref{r1}), with n=1, are
plotted for $\beta=0.7420$,$L_t=13$ (triangles),
$\beta=0.7525$,$L_t=4$ (squares) and
$\beta=0.7585$,$L_t=4$ (crosses). The corresponding  value of the string
tension (as it is listed in tab.II) is subtracted for each sample.
The continuous line is the expected
result if a pure gaussian correction is taken eq.~(\ref{sg}), the dashed
line is the self-avoiding case eq.~(\ref{ssa}). }
\vskip 0.6cm

{\bf Fig.4.}
{\it ~Same as fig.3. The ratios are evaluated at step n=2.
The various samples plotted are:
$\beta=0.7500$,$L_t=10$ (diamonds),
$\beta=0.7500$,$L_t=12$ (circles),
$\beta=0.7525$,$L_t=6$ (squares),
$\beta=0.7585$,$L_t=5$ (crosses),
$\beta=0.7585$,$L_t=6$ (triangles).}

\end{document}